\documentclass[twoside,fleqn]{article}
\usepackage{epsfig,graphicx}
\usepackage{espcrc2}

\title{Larger physical volume with a noncompact lattice regularization 
of SU(N) theories } 
\author{ 
Giuseppe Di Carlo\protect{\thanks{On leave from Laboratori Nazionali di
Frascati, INFN}}
\address{{\small\it Laboratori Nazionali del Gran Sasso -- INFN, 
I-67010 Assergi (L'Aquila), ITALIA.\\
e-mail address: {\tt giuseppe.dicarlo@lngs.infn.it}}},
Fabrizio Palumbo\protect{\thanks{This work has been partially 
  supported by EEC under TMR contract ERB FMRX-CT96-0045}}
  \address{{\small\it INFN -- Laboratori Nazionali di Frascati,
  P.~O.~Box 13, I-00044 Frascati, ITALIA. \\
e-mail address: {\tt palumbof@lnf.infn.it}}}  
and Roberto Scimia\address{{\small\it Marconi Mobile,
Viale dell'Industria, 4 - 00040 Pomezia (Roma), ITALIA 
and Dipartimento di Fisica and INFN -- Sezione di Perugia, 
Universit\`a degli Studi di Perugia
Via A. Pascoli - 06100 Perugia, ITALIA.\\
e-mail address: {\tt roberto.scimia@lnf.infn.it}}}
}

\begin{document}

\newcommand{\lp}{\left}
\newcommand{\rp}{\right}
\newcommand{\be}{\begin{equation}}
\newcommand{\ee}{\end{equation}}
\newcommand{\reff}[1]{(\ref{#1})}
\newcommand{\beqa}{\beqin{eqnarray}}
\newcommand{\eeqa}{\end{eqnarray}}
\newcommand{\ep}{\varepsilon}
\newcommand{\dep}{\partial}
\newcommand{\al}{\alpha}
\newcommand{\ch}{\chi}
\newcommand{\sg}{\sigma}
\newcommand{\tht}{\theta}
\newcommand{\ovr}{\overline}
\newcommand{\sgb}{\overline{\sigma}}
\newcommand{\chb}{\overline{\chi}}
\newcommand{\phb}{\overline{\phi}}
\newcommand{\rhb}{\overline{\rho}}
\newcommand{\cb}{\overline{c}}

\begin{abstract}
Recently it has been found that in a noncompact formulation of the SU(2) 
gauge theory on a lattice the physical volume is
larger than in the Wilson theory with the same 
number of sites.
In its original formulation  such noncompact regularization is directly
applicable to U(N) theories for any N and to SU(N) theories for N=2 only. 
In this work we extend it to SU(N) for any N and
investigate some of its properties.
\end{abstract}

\maketitle

\section{Introduction and summary}
One of the major problems in lattice calculations is the need for a 
large physical volume~\cite{volume}. One possibility is to 
adopt an action for which the scaling is established at a larger lattice 
spacing. In this spirit we 
investigate a noncompact regularization~\cite{Fab} where exact gauge 
invariance is enforced by the use of auxiliary fields.

\section{The  noncompact regularization for U(N)}

As it is well known, a simple discretization of the continuum covariant 
derivative breaks gauge invariance. 
Nevertheless the transformation 
\be
D_\mu \, ' (x)  = g(x) D_\mu (x) g^{\dagger} (x+\mu) 
\ee
of the lattice covariant derivative $D_\mu$ can be exactly enforced by the 
use of auxiliary compensating fields.

Let us assume the following definitions:
\begin{eqnarray}
D_\mu(x) & = &  \left[\frac{1}{a} -\sg_\mu (x) +i\ch_\mu (x) \right] 1\!\!1
\nonumber \\
  &   & + \left[ i A^a_\mu  -\al^a_\mu(x) \right] T_a,\\
F_{\mu \nu}(x) & = & D_{\mu}(x) D_{\nu}(x+\mu)\nonumber \\ 
  &   &            - D_{\nu}(x) D_{\mu}(x+\nu),\\
{\cal L}_{YM}(x) & = & { 1 \over 4} \beta\, \mbox{Tr} 
F_{\mu \nu}^{\dagger}(x)F_{\mu \nu}(x), \label{YM}.
\end{eqnarray}
We emphasize that in such a formulation the measure in the partition
function is flat. 

In the formal continuum limit the field $\sigma_{\mu}$
becomes invariant and decouples together with $\alpha^a_{\mu}$, hence 
these seem to be the auxiliary fields. To control the decoupling of these 
fields at the quantum level we introduce the invariant potential (without 
a continuum analog)
\be
{\cal{L}}_1 = \beta_1 \sum_\mu Tr\,\lp[
D_\mu^{\dagger}(x)D_\mu (x)  -\frac{1\!\!1}{a^2}\rp]^2.
\ee
We obtain an U(N) invariant theory. In general 
we cannot restrict ourselves to the SU(N) symmetry by changing the 
covariant derivative, and at the same time the potential ${\cal{L}}_1$ 
does not generate a mass for the $\chi$-field. Indeed the Ward 
identities~\cite{ncsun} show that no U(N) invariant potential can 
generate a mass for both abelian fields. We must therefore break 
explicitly the U(N) symmetry in order to give all the abelian fields 
a divergent mass. 
The case N=2 is exceptional because by omitting the fields 
$\ch_\mu,\;\al_\mu$ we obtain a SU(2) invariant theory with only one 
abelian auxiliary field. This case has already been exhaustively studied
\cite{BeFab1,FBnp,BeFab2}.

\section{The noncompact regularization for SU(N)}

To break the U(N) invariance of the action we add the potential
\begin{eqnarray}
{\cal{L}}_2 & = & \beta_2\, { 1\over a} \sum_\mu
\lp [  \mbox{det} \,D_\mu(x) + \mbox{det} \,D_\mu^{\dagger}(x) \rp]
\end{eqnarray}
which requires~\cite{ncsun} the additional term 
\begin{eqnarray}
{\cal{L}}_3  & = & \beta_3\, \frac{1}{a^2} \sum_\mu Tr\,\lp[
D_\mu^{\dagger}(x)D_\mu (x)  -\frac{1\!\!1}{a^2}\rp].
\end{eqnarray}
In conclusion the full classical lagrangian is
\be
{\cal{L}}_G={\cal{L}}_{YM}+{\cal{L}}_1+{\cal{L}}_2 +{\cal{L}}_3.
\ee

By adopting a definition of the covariant derivative where the
abelian fields are in a polar representation
\be
D_\mu(x) = \hat{D}_{\mu}(x) \exp\,i \phi_{\mu}(x), \label{polar}
\ee
where
\be
\hat{D}_{\mu} = \rho_{\mu} 1\!\!1 + \left[ i( A')^a_\mu
-(\al ')^a_\mu \right] T_a,
\ee
we find that the lagrangian density is stationary for~$\rhb^{(0)}_\mu = 0$ 
and 
\be
\rhb^{(\pm)}_\mu =
\frac{1}{4a\beta_1}
\lp [|\beta_2|\pm
\sqrt{\beta_2^2+8\beta_1\lp ( 2\beta_1-\beta_3\rp )}\rp ].
\ee
We have to choose the couplings so as to have one and only one 
minimum for $\rhb_\mu = 1/a$; this requirement gives~\cite{ncsun}
\be
|\beta_2|  = \beta_3,\;\; 3\beta_1> \beta_3.
\ee
In the tree approximation the masses of the auxiliary fields turn out to be
\begin{eqnarray}
m^2_{\rho} & = &   { 6\over a^2}\lp(4\beta_1-\beta_3 \rp ),\;\;
m^2_{\phi} = {18\over a^2}\beta_3,\nonumber \\
 m^2_{\al} & = & { 8\over a^2}\lp(2\beta_1+\beta_3 \rp ).
\end{eqnarray}

\section{Numerical study of the SU(2) noncompact regularization}

In the SU(2) case the regularization can be defined with only two 
coupling constants, namely $\beta,\;1/\gamma$. In the limit of 
vanishing $1/\gamma$ we obtain the Wilson regularization.

In perturbative calculations the continuum limit is reached with a scale 
parameter equal to Wilson one~\cite{FBnp}. 

In numerical simulations the situation is different. 
On general grounds we expect that there will exist a scaling region in 
the plane $\beta,\;1/\gamma$ where the properties of convergence to 
the continuum vary. To investigate 
this feature we adopted the following strategy~\cite{GDC-R}. 
The ratio $R(\beta,1/\gamma)$ of two particle masses was calculated 
on a regular grid  in the~$(\beta,1/\gamma)$ plane.  
These values were fitted so as to obtain a 
continuously varying surface. The scaling region is that where the ratio 
is constant and takes the same value it has 
in the asymptotic scaling region. 
Actually in the scaling region the mass ratio must take the same value it 
has in the asymptotic scaling region for Wilson regularization 
(the~$1/\gamma=0$ line). The comparison with Wilson results 
allows us to shorten our calculations, but in no case is mandatory. 
\begin{figure}
\begin{center}
\epsfig{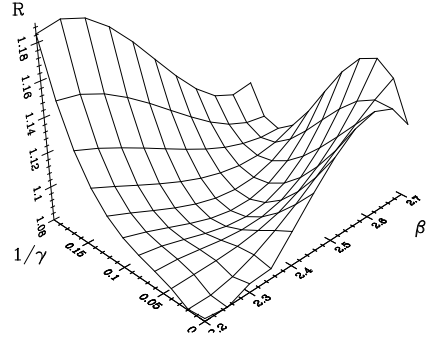}
\caption{$R(\beta,1/\gamma)$ fitting surface.}
\end{center}
\end{figure}

It is a crucial condition for the above scheme to be
valid, that the only dimensionful quantity of the theory be the 
renormalization group scaling parameter. This is true if we work 
in the chiral limit, therefore we evaluated the mass spectrum at four finite 
quark mass values and then we extrapolated it to the chiral limit. 
The same considerations led us to work in the quenched approximation, 
also to allow a comparison with the analytical 
calculations in~\cite{FBnp,BeFab2}.

The details of our numerical work are described in~\cite{GDC-R}. Here we 
briefly discuss some of the results.
In Fig.~1 we report the fitting surface $R(\beta,1/\gamma)$.
We see an (almost) flat region ({\it valley})
that originates from the $1/\gamma=0$ line (Wilson results), and 
propagates towards larger values of $\beta$ for increasing $1/\gamma$. 
The flat region in the Wilson limit coincides with the usual asymptotic 
scaling region for $SU(2)$ pure gauge lattice theory for the small lattices  
we considered.
\begin{figure}
\begin{center}
\epsfig{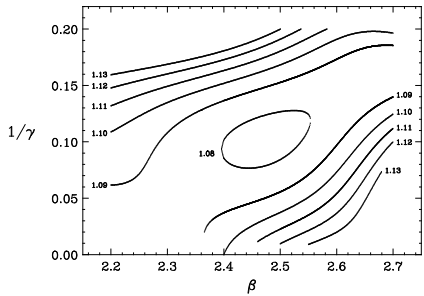}
\caption{Lines of constant \mbox{$R$} in the \mbox{$\beta,1/\gamma$} plane.}
\end{center}
\end{figure}
Therefore we tentatively identify the valley as the scaling
window (although not the asymptotic scaling region) for the non-compact
regularization.
In order to make this observation more precise we report in Fig.~2 the curves 
of constant $R$ in the plane $\beta,1/\gamma$. 
We performed some additional check to confirm the identification 
of the scaling region, obtaining reassuring results~\cite{GDC-R}. 

Any point inside the scaling window is a good one for approximating
the continuum theory, but actual results can be different. 
In particular the physical value of the lattice spacing, and then the 
physical volume of the lattice, varies from point to point. 
\begin{figure}
\begin{center}
\epsfig{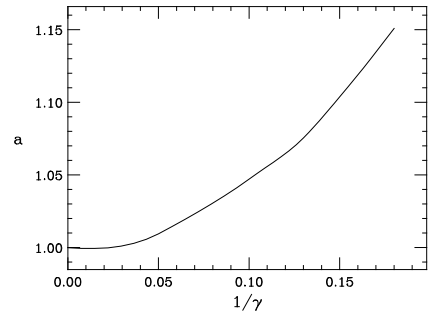}
\caption{Ratio of non-compact and Wilson lattice spacing along a
curve lying on the bottom of the scaling valley. }
\end{center}
\end{figure}
We present in Fig.~3 the behaviour of the lattice spacing, as extracted
from the $\rho^+$ mass, along the {\it center} of the valley.
We can clearly see that the lattice spacing becomes larger and larger 
the more we depart from the Wilson case, $1/\gamma=0$.
This improvement can be increased if we move towards
larger $\beta$ and $1/\gamma$, but with a narrower scaling valley.


\begin{thebibliography}{9}

\bibitem{volume} R. Petronzio, Nucl. Phys. B (Proc. Suppl.)
{\bf 83-84} (2000) 136; D.K.Sinclair, Nucl. Phys. B (Proc. Suppl.) {\bf 47}
(1996) 112; L.Lellouch, Nucl. Phys. B (Proc. Suppl.) {\bf 94} (2001) 142.

\bibitem{Fab}
F. Palumbo, Phys. Lett. {\bf B244}, (1990) 55.

\bibitem{ncsun}
F. Palumbo and R. Scimia, hep-lat/0105029

\bibitem{BeFab1}
C. M. Becchi and F. Palumbo, Phys. Rev. {\bf D44} (1991) R946.

\bibitem{FBnp}
C. M. Becchi and F. Palumbo, Nucl. Phys. {\bf B388} (1992) 595.

\bibitem{BeFab2}
F. Palumbo, M. I. Polikarpov and A. I. Veselov, Phys. Lett.
{\bf B297} (1992) 171;
B. Diekmann, D. Schutte, and H. Kroger, Phys. Rev. {\bf D49}
(1994) 3589;
B. Borasoy, W. Kramer and D. Schutte, Phys. Rev.
{\bf D53} (1996) 2599.

\bibitem{GDC-R}
G. Di Carlo and R. Scimia, Phys. Rev. {\bf D63} (2001) 094501 ,
 hep-lat/0009019.

\end{thebibliography}
\end{document}